\documentclass{jps-cp}

\title{Dark matter and Dark photons in formal theory}

\author{Gennady \textsc{Kozlov} }

\inst{Joint Institute for Nuclear Research, Joliot-Curie st. 6, Dubna, 141980, Russia }

\email{kozlov@jinr.ru}

\abst{ We find the Standard Model sector (SM) interacts to Dark matter (DM) sector  through the scale invariant degrees of freedom - the dilatons and Dark photons (DP). The latter have an interplay with SM via the derivatives of the scalar dilaton fields. }

\kword{Dark matter, Dark photon, dilaton.}

\begin{document}
\maketitle

\section{Introduction}

Despite the efforts for direct or indirect search, Dark matter (DM) and in particular Dark photons (DP) still are in the hidden sector of matter. The difficulty so far is that DP have been probed only in terms of Standard Model (SM). In particular, the restrictions for DP (the mass, the kinetic mixing angle with SM photon) are given by its decays to charged leptons, including neutrinos within missing energy.  DP may be one of the forms of DM where the density of the latter is made of the energy density of the oscillations of Dark photons [1].
There is a growing interest to thermal freeze-out weakly interacting candidates for DM $\chi$ in the mass range $m_{\chi} \sim keV - GeV$. If the dominant freeze-in process is an annihilation of SM particles into DM via a light vector mediator, e.g., DP with mass $m < m_{\chi}$, the thermal cross-section typically scales as $\langle\sigma\, v\rangle \sim g^{2}\,g^{2}_{SM} / (4\pi T)^{2}$, where $g$ is the coupling of DP to DM, $g_{SM}$ is the coupling of DP to SM, and $T$ is the temperature of the bath. 

The physics of DP has been widely studied in the literature (see, e.g., [2] and the refs. therein).  DP as a hidden light gauge boson plays an important role in the understanding of electroweak baryogenesis where CP violation occurs in dark sector [3].
However, to the author's opinion, the whole matter has not been settled in field theory in a satisfactory way. In particular, the very definition of DP field as a solution of respected equation of motion is still lacking. It is important to look at the nature of DP  through the first principles of quantum field theory.

It is known the interest to  theories in which SM is strongly coupled to a conformal sector of particles. In the early Universe or at early times of heavy ion collisions, the quantum fluctuations appeared can produce a strong background of scalar dilaton field due to an explicit breaking of conformal (scale) symmetry. The approximate scale-invariant sector includes both DM and SM fields. Since DM has no direct couplings to SM, the dilaton is the background of the messenger between SM and DP, and further DM. DP plays the role of mediator between DM and SM.
The DM and DP thermal relic abundance are governed by their couplings to the dilaton, and thus, strictly related to breaking scale $f$ for fixed masses of DM and DP.
Because of conformal anomaly where dilatation current $S_{\mu}$ is not conserved ($\partial^{\mu} S_{\mu} (x) \neq 0$), the dilaton  may decay to massless or massive vector states, photons or DP's. The cross section of Dark photons $(\bar\gamma \bar\gamma)$ production is $\sigma (\Phi\rightarrow\bar\gamma \bar\gamma) = \sigma (h\rightarrow \gamma \gamma)\cdot CR_{\gamma \bar\gamma}$ with
$$ CR_{\gamma \bar\gamma} = \frac{\Gamma_{\Phi\rightarrow \bar\chi\chi} \cdot \Gamma_{\Phi\rightarrow\bar\gamma \bar\gamma}/\Gamma^{tot}_{\Phi}}{\Gamma_{h\rightarrow gg} \cdot \Gamma_{h\rightarrow \gamma \gamma}/\Gamma^{tot}_{h}},$$
where dilatons $\Phi$ are produced via DM $\bar\chi \chi$ annihilation. The light hadron $h$ is the subject of gluon-gluon $(gg)$ fusion and the decay to two photons $(\gamma \gamma)$.
The dilaton differs from other spin-0 particles by its couplings to massless gauge bosons.
For applications  to accelerator experiments (both for collider's and fixed target modes), the couplings of dilaton $\Phi$ with gauge bosons at energies below $4\pi f$ are given by trace of the energy-momentum tensor $\theta_{\mu}^{\mu}$:
$$ \theta_{\mu}^{\mu} = \partial^{\mu} S_{\mu} = \frac{\alpha_{EM}}{8\pi f} c_{EM}\,\Phi \, (F_{\mu\nu})^{2} + \frac{\alpha_{s}}{8\pi f} c_{g}\,\Phi \, (G_{\mu\nu}^{a})^{2} + \frac{\alpha_{D}}{8\pi f} c_{D}\varepsilon^{2}\,\Phi \, (B_{\mu\nu})^{2}, $$
where $f = \langle \Phi\rangle$ is the order parameter for scale symmetry breaking determined by the dynamics of the underlying strong sector; $\alpha_{EM} (c_{EM}), \alpha_{s} (c_{g})$ and $\alpha_{D} (c_{D})$ are the couplings (coefficients) relevant to electromagnetic, strong and DM sectors; $F_{\mu\nu} = \partial_{\mu} A_{\nu} - \partial_{\nu} A_{\mu}$ with photon field $A_{\mu} (x)$,    $B_{\mu\nu} = \partial_{\mu} B_{\nu} - \partial_{\nu} B_{\mu}$ with DP field $B_{\mu} (x)$, and $G_{\mu\nu}^{a}$ is the canonically normalised gluon strength tensor; $\varepsilon$ is the kinetic mixing between the photon and DP.  
Depending on the implementation of the model,  the strength $\varepsilon$ can range from  $10^{-12}$ to $10^{-2}$. In the low energy experiments, the  values of $\varepsilon$ in the window $10^{-7} - 10^{-3}$ have been probed [2]. If no excess events are found, the obtained results can be used to impose bounds on the  mixing strength $\varepsilon$ as a function of DP mass $m$. The mass  $ m = 0.83 GeV$, and $\varepsilon = 7.6\cdot 10^{-3}$ were estimated in [4] with electromagnetic neutrino form-factor calculations. 
The results with the dependence of  $\varepsilon$ should be compared to the data of transverse momenta spectra of direct (thermal) photons.
Since the SM is embedded in conformal sector one can use the conformal invariance condition 
\begin{equation}
\label{e1}
\sum_{light} b_{i} +   \sum_{heavy} b_{i} = 0 ,
\end{equation}
where $i$ carries either QCD or electroweak (EW) features of the coefficients $b_{i}$ of corresponding $\beta$-functions.
The sum in (\ref{e1}) is splitted over all coloured particles into sums over light and heavy states in the mass scale separated by  mass of the dilaton. 
Hence the only quarks lighter than that of the dilaton are included in the coefficients of corresponding $\beta$-function. 
In particular, $b^{light}_{QCD} = - c_{g} = -11 + 2n_{light}/3$ in the interaction between the dilaton and gluons,   where the number of light quarks is either  $n_{light} = 5$ if the mass of the dilaton $\mu < m_{t}$, or $n_{light} = 6$ if $\mu > m_{t}$ for the top quark mass $m_{t}$ [5]. In EW sector one has $c_{EW} = -80/9$ if $\mu < 2 m_{W}$, or $c_{EW} = -35/9$ if $2m_{W} < \mu < 2 m_{t}$, or $c_{EW} = -17/3$ if $2 m_{t} < \mu$ [6], where  $m_{W}$ is the mass of the $W$-boson. 

The certain gauge models may admit the additional $U^{\prime}(1)$ Abelian symmetry for which SM fields carry no charge. Such a gauge group is associated  with new  gauge bosons  which can have small masses or will be almost massless (for a  review see [7] and the references therein).
In these models, the extended group $SU(2)_{L}\times U(1)_{Y}\times U^{\prime}(1)_{B}$ may appear, where  index $B$ in  $U^{\prime} (1)_{B}$ is associated with a hidden  photon  $\bar\gamma$ given by the gauge field $B_{\mu}$. The standard photon $\gamma$ may oscillate into  $\bar\gamma$  where the latter being the short-lived state can decay to invisible neutrino-antineutrino pair, $\bar\gamma\rightarrow \bar\nu\,\nu$ or to electron-positron pair $\bar\gamma\rightarrow e^{-}e^{+}$. 

The dilaton gets the vacuum expectation value (vev) $f$ when the conformal invariance is breaking down. The latter triggers electroweak symmetry breaking  at the scale $\Lambda_{EW} = 4\pi v < \Lambda_{CFT} = 4\pi f$, where $v =246$ GeV is the vev of the Higgs boson. The scales $f$ and $v$ are different except for the Higgs boson ($f = v$). If the approximate conformal symmetry is broken at $\Lambda_{CFT}$, the low energy spectrum of composite states may contain a light dilaton, a light Higgs doublet or both.
The dilaton operator triggers the breaking of  $SU(2)_{L}\times U(1)_{Y}$ gauge invariance through  the dilaton mass operator. 
The scalar colour-singlet state $\Phi (p)$  with the mass $\mu$,  the momentum $p_{\mu}$ and the decay constant $f$ can be produced when dilatation current $S_{\mu}$ acts on the vacuum
$ \langle \Omega\vert S^{\mu} (x)\vert \Phi(p)\rangle = i\,p^{\mu}\,f\,e^{-i\,p\,x},\,\,\, \langle \Omega\vert \theta^{\mu}_{\mu}(x)\vert \Phi(p)\rangle = \mu^{2}\,f\,e^{-i\,p\,x},$ 
where $\vert \Omega\rangle$ is the vacuum state corresponding to spontaneously broken dilatation symmetry.

This letter is an attempt to clarify the properties of DM, DP and dilatons using the Lagrangian approach with the canonical quantisation.  The whole subject is examined in this letter from the beginning by studied the Abelian Dipole Dark Photon Model (ADDPM). We present a soluble four-dimensional ADDPM which has the formal similarities to the Abelian Higgs model. The model exhibits a $\delta^{\prime} (p^{2})$ singularity that characterises the two-point function of the dipole field, in our case the scalar dilaton field obeying the equation of the 4th order, in gauge invariant field theory.
The solution for DP contains the derivatives of the scalar dilaton field.

\section{Dilaton and Higgs boson}
Let us consider the Lagrangian density (LD) of the classical scalar field $\Sigma$ for two particles with complex conjugate masses $s$ and $s^{\star}$
$$ L_{cl} = \frac{1}{2i}\left [\left (\partial_{\mu}\Sigma\right )^{2} - s^{2}\,\Sigma^{2} \right ] -  \frac{1}{2i}\left [\left (\partial_{\mu}\Sigma^{\star}\right )^{2} - s^{\star\,2}\,\Sigma^{\star\, 2} \right ].$$
The transformations $\Sigma = (H+i\,D)/\sqrt {2}$, $\Sigma^{\star} = (H-i\,D)/\sqrt {2}$ lead to LD (see also [8])
$$ L (x) = \partial_{\mu} H (x)\partial^{\mu} D (x) - \nu^{2}\,H(x)\,D(x) -\frac{1}{2} \,m_{H}^{2}\,H^{2} (x),$$
where $H$ and $ D$  are the Higgs boson field and the dilaton field, respectively; $\nu$ is the mixing mass and $m_{H}$ is the mass of $H$. We have the following equations of motion 
$(\Box + \nu^2) H(x) = 0$, $(\Box + \nu^2) D(x) = - m_{H}^{2} H (x)$ accompanied by commutation relations 
$$ [H(x), H(x^{\prime}] = 0,  [H (x), D (x^{\prime})] = i\Delta (x - x^{\prime};\nu^{2}),  [D(x), D(x^{\prime})]  = i\,m_{H}^{2} \frac{\partial}{\partial \nu^{2}}\, [\Delta (x- x^{\prime}; \nu^{2})],$$ 
where
$$i \Delta (x - x^{\prime}; \nu^{2}) = \int d_{4} k e ^{-i k (x- x^{\prime})} [\tilde H (k), \tilde D (k^{\prime})] \delta (k^{2} - \nu^{2}) = \int d_{4} k e^{- i k (x - x^{\prime})} \epsilon (k^{0}) \delta (k + k^{\prime}).$$ 
The dilaton field is decomposed as $D(x) = D_{1} (x) + D_{2} (x)$, where 
$$D_{1} (x) = \frac{m_{H}^{2}}{\nu^{2}}\, H(x) + C(x), \,\,\,D_{2} (x) = \frac{m_{H}^{2}}{2\nu^{2}}\, x_{\mu}\, \frac{\partial} {\partial x_{\mu}} H(x) $$
 with $(\Box + \nu^{2} ) C(x) = 0.$  The field $D_{2}(x)$ is the dipole ("ghost") part of the dilaton field obeying the equation $ (\Box + \nu^{2})^{2} D_{2} (x) = 0.$ Our model developed in Sections 3 and 4 has some resemblance to soluble model [9] in the limit $\nu^{2}\rightarrow 0$.
 
\section{Abelian Dipole Dark Photon Model}
We assume the approximate scale invariance of ADDPM in order to build an effective LD for energies below $\sim 4\pi f$ where the scale invariance is preserved by the dilaton field $\Phi$. In the presence of the source, LD is 
\begin{equation}
\label{e2}
L = -\frac{1}{4} F_{\mu\nu}^{2} - \frac{1}{2}{\varepsilon} F_{\mu\nu}B^{\mu\nu} - \frac{1}{4}{\varepsilon^{2}} B_{\mu\nu}^{2} + \bar\chi (i\hat D - m_{\chi})\chi + \vert D_{\mu_{1}}\Phi\vert^{2} - \lambda \vert \Phi\vert^{4} + \mu_{0}^{2}\vert  \Phi\vert^{2} - b (\partial\cdot B) + \frac{b^{2}}{2\eta},
\end{equation}
where $ D_{\mu} = \partial_{\mu} + i g B_{\mu}$, $\hat D = D_{\mu}\gamma^{\mu}$; $D_{\mu_{1}} = \partial_{\mu} + i g_{1} B_{\mu}$; $\chi$ is the Dirac DM field with mass $m_{\chi}$; $\lambda$ is the self-coupling constant of the dilaton field; $\mu_{0}$ and $\eta$ are real parameters. The mixing of a photon and DP is  induced by the shift $A_{\mu} \rightarrow A_{\mu} + \varepsilon B_{\mu}$, where $\varepsilon$ is free parameter (mixing angle). The DM field $\chi$ is neutral under SM quantum numbers and charged under a hidden gauge symmetry that is broken at low energies. LD (\ref{e2}) is invariant under the restricted gauge transformations
$$ A_{\mu}\rightarrow A_{\mu} + \partial_{\mu}\Lambda,\,\, B_{\mu}\rightarrow B_{\mu} + \partial_{\mu}\Lambda, \,\,
 \Phi\rightarrow \Phi e^{-ig_{1}\Lambda}, \,\chi\rightarrow \chi e^{ig\Lambda},\,\, b\rightarrow b,$$
where $\Lambda (x)$ satisfies $\Box\Lambda (x) = 0$. The field $b(x)$ plays the role of gauge-fixing multiplier in LD (\ref{e2}), and it remains free. To proceed to the solution of ADDPM, we consider the real scalar fields
\begin{equation}
\label{e3}
\phi + f = \frac{1}{\sqrt 2} (\Phi + \Phi^{\star}),\,\,\, \varphi  = \frac{-i}{\sqrt 2} (\Phi - \Phi^{\star}),
\end{equation}
where 
\begin{equation}
\label{e4}
\langle\Omega,\varphi \Omega\rangle = 0, \,\,\, f = \langle\Omega, (\phi + f) \Omega\rangle, \,\, \langle\Omega, \Omega\rangle = 1.
\end{equation}
Having in mind (\ref{e3}) and (\ref{e4}), LD (\ref{e2}) becomes:
\begin{equation}
\label{e5}
L = L_{1} + L_{2},
\end{equation}
where (the zeroth order of $g_{1}f$ and $\lambda f$ are considered)
\begin{equation}
\label{e6}
L_{1}  = -\frac{1}{4} F_{\mu\nu}^{2} - \frac {1}{2}{\varepsilon} F_{\mu\nu}B^{\mu\nu} - \frac {1}{4}{\varepsilon^{2}} B_{\mu\nu}^{2} + \bar\chi (i\partial_{\mu}\gamma^{\mu} - m_{\chi})\chi 
- g(B\cdot J) - b (\partial\cdot B) + \frac{1}{2\eta} b^{2},
\end{equation}
\begin{equation}
\label{e7}
L_{2}  =  \frac{1}{2}{m^{2}} B_{\mu}^{2} + m B_{\mu}\partial^{\mu}\varphi - \frac{1}{2}\mu^{2} \phi^{2} + 
\frac{1}{2} \left [\left (\partial_{\mu}\phi\right )^{2} + \left (\partial_{\mu}\varphi\right )^{2}\right ].
\end{equation}
In Eqs. (\ref{e6}) and (\ref{e7}), $J_{\mu} = \bar\chi\gamma_{\mu}\chi$ is the current of DM, $ m = g_{1} f$ is the DP mass, $\mu = \sqrt{2}\lambda f$ is the dilaton mass. The equations of motion are:
\begin{equation}
\label{e9}
m^{2} B_{\mu} + \eta\partial_{\mu} (\partial\cdot B) + m\partial_{\mu}\varphi - g J_{\mu} = 0,
\end{equation}
\begin{equation}
\label{e11}
\Box^{2} \varphi = 0,\,\,\, \Box\varphi \neq 0.
\end{equation}
The solution of Eq. (\ref{e9}) is
\begin{equation}
\label{e12}
B_{\mu} = \frac{g}{m^{2}} J_{\mu} - \frac{1}{m} \partial_{\mu}\varphi + \frac{\eta}{m^{3}} \partial_{\mu}\Box\varphi.
\end{equation}

\section{Propagators}
We introduce two-point Wightman function (TPWF) $W(x) = \langle \Omega, \varphi (x) \varphi (0)\Omega\rangle$ which obeys the equation
 \begin{equation}
\label{e13}
\Box^{2} W(x) =0
\end{equation}
taking into account (\ref{e11}).  The general solution of (\ref{e13}) is (see [10] and the refs. therein)
\begin{equation}
\label{e14}
W(x) = a_{1} \,\ln \frac {l^{2}} {-x_{\mu}^{2} + i\epsilon x^{0}} + a_{2}\, \frac{1}{x_{\mu}^{2} - i\epsilon x^{0}} + a_{3},
\end{equation}
which is the distribution on the space $S^{\prime}(\Re ^{4})$ of temperate generalised function on $\Re ^{4}$. The space $S^{\prime}(\Re ^{4})$ is conjugate to the (complex) space  $S (\Re ^{4})$ to the test functions on $\Re ^{4}$.  
The coefficients $a_{1}$ and  $a_{2}$ in (\ref{e14}) will be defined later, while $a_{3}$ is an arbitrary constant.
The parameter $l$ in (\ref{e14}), having the dimension in units of length,  breaks the scale invariance under dilatation transformation.
The TPWF (\ref{e14}) is the homogeneous generalised function of the zeroth order with the dilatation properties $W (\rho x) = W (x) - 1/(8\,\pi)^{2}\,\ln\rho,\,\,\,\rho > 0$.

Since the TPWF of dilaton field does not admit the Kallen-Lehmann representation, $\varphi (x)$  is the quantum field defined in the space with an indefinite metric. We consider this statement more carefully.
The Fourier transformation (FT) of the first term in (\ref{e14}) is given by  
$$\int 2\,\pi\,\theta (p^{0})\,\delta^{\prime}(p^{2}, \hat M^{2})\,e^ {-ipx} d_{4}p, $$ where $\hat M = (2/l)\, e^ {1/2-\gamma}$, $\gamma $ is the Euler's constant. 
Here, we deal with the product $ \theta (p^{0})\,\delta^{\prime}(p^{2}, \hat M^{2})$ to generalised functions. This product is well-defined distribution only on the space $S(\Re_{4})$ of complex Schwartz test functions $u(p)$,
and would have the form as $\delta^{\prime} (p^{2})$ only on the space  $S_{0}(\Re_{4})$, where the test functions $u(p)$ from $S(\Re_{4})$ are zero at $p = 0$. 
The Hermitian form  $\langle\Omega, \varphi (\tilde f)\, \varphi (\tilde g)\Omega\rangle$ ($\tilde f, \tilde g \in  S (\Re _{4}))$ on $S (\Re _{4})$  does not defined as the positive one. The reason of the latter is followed from  dilatation properties of $W (\rho x)$ as well as from 
$$\int 2\,\pi\,\,\tilde f (p)\,\theta (p^{0})\,\delta^{\prime}(p^{2})\,\tilde g (p)\, d_{4} p = \int_{\Gamma_{0}^{+}} \, \frac{1}{2\,n\,p} \,\left ( -n\,\partial + \frac{1}{n\,p}\right )\,\tilde f (p) \,\tilde g (p) \, \frac{d^{3} p}{2p^{0}},$$
where $n$ is the fixed unit time-like vector ($n^{2} =1 $) in the Minkovsky space $\mathcal {S} (\mathbb {M})$ from $V^{+} = \{p \in\mathbb {M}, p^{2} > 0, p^{0} > 0\}$; $\Gamma_{0}^{+} = \{p \in \mathbb {M}, p^{2} = 0, p^{0} > 0\}$, $n\partial = n_{\mu}\,\partial/{\partial p_{\mu}}$. Therefore,  $\theta (p^{0})\,\delta^{\prime} (p^{2})$ is defined as the distribution from $S^{\prime}(\Re ^{4})$ through the first term in  (\ref{e14}). 
For $n = (1, \vec {0})$ one has
$$- \frac{\partial}{\partial \mu^{2}} \int 2\,\pi\,\theta (p^{0})\,\delta(p^{2} - \mu)^{2})\,u (p)\, d_{4} p = - \frac{\partial}{\partial \mu^{2}} \int \frac{d_{3} p}{2\,E} \,u (E, \vec {p}). $$
What we find out is the presence of $\delta^{\prime} (p^{2}, \hat M^{2})$ in FT of 
$W (x)$ is a consequence of the nonunitarity of translations ($\delta^{\prime} (p^{2}, \hat M^{2})$ is not a measure). 
Hence, the secondary quantised formalism relevant to the field $\varphi (x)$ has to be built up in the space with indefinite metric.
 The latter is useful to the formalisation of the idea of virtual states. For this, the Hilbert space of physical state vectors has to be replaced by pseudo-Hilbert space of virtual state vectors (see Sec. 6).

The commutator for $\varphi (x)$ field is
$$ [\varphi (x), \varphi (0)] = i\eta^{-1} \left [ D(x) - m^{2} E(x) \right ], $$
where 
$$ D (x) = (2\pi)^{-1}\epsilon (x^{0})\,\delta(x^{2}), \,E (x) = (8\pi)^{-1}\epsilon (x^{0})\,\theta(x^{2}) $$
with the following properties:  $\Box D(x) = 0$, $\Box E (x) = D(x)$, $\Box^{2} E (x) = 0$. The coefficients $a_{1}$ and $a_{2}$ from (\ref{e14}) are taken into account. In particular, 
$a_{1} = -m^{2}/(16\pi{^2}\eta)$ is obtained from the canonical commutation relation 
 $ \left [\varphi(x), \pi(\varphi(0))\right ]_{\vert x^{0} = 0} =i\,\delta^{3} (\vec x) $,
where $\pi (\varphi) = m B_{0} + \partial_{0}\varphi $ is the conjugate momentum to $\varphi$. The coefficient $a_{2} = 1/(4\pi^{2}\eta)$ is found from the dimensional analysis.

The commutators for DP field $B_{\mu}$ follow from the solution (\ref{e12}) for all $x$ and $y$:
$$\left [B_{\mu} (x), \varphi (y)\right ] = \frac{i}{\eta\, m} \partial_{\mu} \left [ m^{2} + (\eta -1)\Box \right ] E(x-y),$$
$$\left [B_{\mu} (x), b (y)\right ] = \frac{i}{\eta \,m^{2}} \partial_{\mu}  D(x-y),\,\,\,\,\,
\left [B_{\mu} (x), B_{\nu}(y)\right ] = \frac{-i}{\eta} \partial_{\mu}\partial_{\nu}  E(x-y).$$
The commutators for $b(x)$ are
$$\left [b(x), \varphi (y)\right ] =  i\, m D(x-y),\,\,\, \left [b(x), b (y)\right ] = 0.$$ 
Finally, we find the commutators with DM field $\chi (x)$. From $\partial_{0} b = g J_{0} - m\partial_{0}\varphi - m^{2} B_{0}$ one finds $[\partial_{0} b(t, \vec x), \chi (t, \vec y) ] = g\,\delta^{3} (\vec x - \vec y)\,\chi (t, \vec y)$ which with $\partial_{0} D (0, \vec x) = \delta^{3} (\vec x)$ gives 
$$    \left [b (x), \chi (y)\right ] =  g  D(x-y) \chi (y). $$
Similarly, we find
$$    \left [\varphi (x), \chi (y)\right ] = -\frac {m}{\eta} g  E(x-y) \chi (y), $$
and hence also
$$    \left [B_{\mu} (x), \chi (y)\right ] =  g\partial_{\mu} \left [\eta^{-1} \,E (x-y) - m^{-2}\,  D(x-y)\right ] \chi (y). $$
In case of weak coupling $g <<1 $, one can get the DM field in the form
$$\chi (x) = \chi_{0} (x): \exp \{-ig[1-(\eta/m^{2})\Box]\varphi (x)/m\}:,$$
where $\chi_{0} (x)$ is the canonical free Dirac field of DM that commutes with $\varphi (x)$. Here,
$$  : e^{-ig\tilde\varphi} : = e^{-ig\tilde\varphi^{(+)}}  \, e^{-ig\tilde\varphi^{(-)}} , $$
where $\tilde\varphi (x) \simeq [1- (\eta/m^{2})\Box ]\varphi (x)/m $, $ \tilde\varphi (x) =   \tilde\varphi^{(-)} (x) +  \tilde\varphi^{(+)} (x)$, $\tilde\varphi^{(+)} (x) = [\tilde\varphi^{(-)} (x)]^{+}$, $\tilde\varphi^{(-)} (x)\Omega = 0$. 

The propagator of DM is calculated at large distances and has the form
$$G_{DM}(x) = \langle \Omega, T [\chi (x) \bar\chi (0)]\Omega\rangle \simeq \frac{1}{(-\kappa^{2} x_{\mu}^{2} + i\epsilon)^{-\alpha_{DM}/(4\pi\eta)}}\, \langle \Omega, T [\chi_{0} (x) \bar\chi_{0} (0)]\Omega\rangle, $$
where DM field $\chi(x)$ has picked up an anomalous dimension $\alpha_{DM} /(4\pi\eta)$,  $\alpha_{DM} =g^{2}/(4\pi)$, and $\kappa\sim l^{-1}$ is the multiplicative normalisation constant of $\chi$.

The propagator of dilaton field in $\Re^{4}$ is
\begin{equation}
\label{e177}
 G_{\varphi} (x) = \langle \Omega, T [\varphi (x) \varphi (0) ]\Omega\rangle = \frac{-m^{2}}{\eta (4\pi)^{2}} \left [ \ln \vert \kappa^{2} x_{\mu}^{2}\vert + i\pi\theta(x^{2}) \right ] + \frac{1}{(2\pi)^{2}} \left [ \frac{1}{x_{\mu}^{2}} + i\pi\delta (x^{2})\right ] + a_{3}, 
 \end{equation}
and its Fourier form in $\Re_{4}$ looks like
\begin{equation}
\label{e1777}
i \hat G_{\varphi} (p) = \frac{m^{2}}{\eta} \left [ \frac{1}{(p^{2} + i\epsilon)^{2}} + i\pi^{2}\ln\epsilon\,\delta^{4} (p)\right ] - \frac{1}{p^{2} + i\epsilon}.
\end{equation}
The coefficient $a_{3}$  in (\ref{e177}) is fixed in such a way as to cancel the constant terms in the expansion of Bessel function $K_{0} (z)$ at small argument $z$ originated from Fourier transformation of $(p^{2} + i\epsilon)^{-2}$ term in (\ref{e1777}). 
We also give the propagator of DP field $B_{\mu}(x)$ in $\Re_{4}$:
$$ i \hat G_{{\mu\nu}}(p) = \langle \Omega, T [B_{\mu} (x) B_{\nu} (0) ]\Omega \rangle = \frac{1}{\eta} \frac{p_{\mu} p_{\nu}}{(p^{2} + i\epsilon)^{2}} - \left ( g_{\mu\nu} - \frac{p_{\mu} p_{\nu}}{p^{2} + i\epsilon} \right ) \frac{1}{p^{2} - m^{2} + i\epsilon} .$$

\section{Observables}
Within the theories of dynamical systems with constraints one can admit  the surface $ b \simeq 0$ in the physical subspace of the phase space corresponding to  (\ref{e5}). According to paper by L. Faddeev [11], any observable is characterised by those functions, the changing of which do not depend with time on the arbitrary functions chosen, e.g., $b(x)$. It means, neither DP $B_{\mu} (x)$, nor DM $\chi (x)$ are not observables because their Poisson brackets $ \{b(x), B_{\mu} (y) \} \sim \partial_{\mu} D (x-y)$ and   $\{b(x), \chi (y)\} \sim g D (x-y)$ are not equal to zero. 

Both DM and DP fields are observables only under stochastic (random) forces represented here by the operator $h_{\mu}$. DP field $B_{\mu} (x)$ has the fluctuations in medium with the probability $\Pi [B_{\mu}] \sim \exp ( - Z [h_{\mu}] )$, where 
$$ Z [ h_{\mu}] = \ln \int d B_{\mu} \exp \left \{ {\int d x [ L(x) + h_{\mu} ( x) B^{\mu} (x) ]} \right \}, $$
and $ h_{\mu}$ is a tempered distribution satisfying $\partial_{\mu} h^{\mu} (x) = \delta (x)$. $ Z[h_{\mu}]$ is entered the free energy $ F_{h}$ which is 
$$ F_{h} = \int d h_{\mu} \, Z [h^{\mu}] \exp \left \{{-\int dx h_{\mu}^{n} (x)}\right \}, $$
$ n = 1,2, ....$ are external insertions of $h_{\mu} (x)$.  The following fields for DM and DP 
$$\tilde\chi (x;h) = \left\{\exp\left [i\,g\int d^{4} y \,h_{\mu} (x-y) \,B^{\mu} (y) \right ]\right \} \chi (x),$$
$$\tilde B^{\mu} (x;h) = \int d^{4} y \left [ g^{\mu\nu} \,\delta (x-y) - \partial ^{\mu} h^{\nu} (x-y) \right ] B_{\nu} (y)$$
obey the following relations in Poisson brackets, respectively, $\{b(x), \tilde\chi (y;h)\} = 0,$ $ \{b(x), \tilde B_{\mu} (y;h)\} = 0, $
and, hence, $\tilde\chi$ and $\tilde B_{\mu}$ are observables. Moreover, the latter are local if $ \{O(x;h), O (y;h)\} = 0$ for all $(x-y)^{2} < 0$, where $O: \tilde\chi,\, \tilde B_{\mu}$.

\section{Physical states}
In section 3 we found $\varphi (x)$ is a field defined in the space  with an indefinite metric which can not be clarified in terms of physical states. One can provide these states would not appear in the asymptotic expressions for physical observables at  $ t \rightarrow \pm\infty$. We follow after Heisenberg [12] and Bogolyubov [13] in $S$-matrix theory. Let us consider the local scalar fields $\Phi (x)$  in the form 
$$\Phi (x) = z(x) + \sum_{n} c_{n}\,\zeta_{n} (x), \,\,\, c_{n} = const,$$
where some of fields $\zeta_{n}(x)\in \{\zeta_{1} (x), ..., \varphi (x), ...\}$ may have the commutation relations with negative sign; $z (x)$ is the set of real (physical) states, e.g., Higgs-boson among them; the dilaton field $\varphi (x)$ may stand as a virtual (fictitious) state. We introduce  two Hilbert spaces: the real space $\ S (\Re^{4})$ with respect to real (physical) particles describing by $z(x)$, and  the space $\ S^{\prime} (\Re^{4})$ for $\zeta_{n} (x)$ fields. The total Hilbert space is $\mho (\Re ^{4}) =  S (\Re^{4}) +  S^{\prime} (\Re^{4})$. 

We suppose that each amplitude of the state possesses by both physical and virtual parts, however the part of the amplitude corresponding to $\zeta$ state is defined unique by its physical part based on the state $z$. 
In the operator form $\Phi$ is divided into two parts $\Phi = z + \zeta$, where $z = P\,\Phi$ ($z \in S(\Re ^{4})$) and $\zeta = (1 - P)\,\Phi$ ($\zeta \in S^{\prime}(\Re ^{4})$). Here, $P$ is the operator which projects the states $\Phi$ from $\mho (\Re^{4})$ to $S (\Re ^{4})$; $P ^{+} = P$, $ P^2 = P$; ${\parallel\Phi\parallel}^2 = {\parallel z \parallel}^2 + {\parallel\zeta\parallel}^2$, ${\parallel z \parallel}^2 > 0$. In the $S$-matrix approach $\Phi_{+ \infty} = S \,\Phi_{- \infty}$. 
We suppose that the system is found in some state $\Phi_{-\infty} = z_{-\infty} + \zeta_{-\infty}$ at $t\rightarrow - \infty$ first, and because of interactions, the system turns into the state $\Phi_{+\infty} = z_{+\infty} + \zeta_{+\infty}$ at $t\rightarrow + \infty$. The following condition $\parallel\Phi_{-\infty}\parallel = \parallel\Phi_{+\infty}\parallel$ is evident.
 Then one finds 
\begin{equation}
\label{e44}
z_{+ \infty} = P\,S (z_{- \infty} + \zeta_{- \infty}),
\end{equation}
\begin{equation}
\label{e45}
\{ \zeta_{- \infty} + (1 - P)\,S (z_{- \infty} + \zeta _{-\infty})\} = 0,
\end{equation}
where the nonlocal boundary condition $\zeta _{- \infty} + e^{i\,\delta}\,\zeta _{+ \infty} = 0$ is used ($\delta$ is a phase).  Eq. (\ref{e45}) defines the unphysical (virtual) part of the amplitude at $t\rightarrow -\infty$ by the physical part of the amplitude
$$\zeta_{-\infty} = - \{1 + (1 - P)\,S\}^{-1}\, (1 - P)\,S\, z_{-\infty}.$$
The input physical states of the system is described by the vector states of the form:
$$\Phi_{-\infty} = z_{-\infty}  - \{1 + (1 - P)\,S\}^{-1}\, (1 - P)\,S\, z_{-\infty}.$$
From (\ref{e44}) and (\ref{e45}) one finds the asymptotic virtual state  $\zeta _{+\infty}$ through the asymptotic state $\ z_{-\infty}$ of real particles
 $$\zeta_{+ \infty} = \{ 1 + (1 - P)\,S \}^{-1} (1- P)\,S\, z_{- \infty}. $$
Here,  $z_{- \infty}$ is defined through the equation $z_{+\infty} = \tilde S\, z_{-\infty}$ and the unitary matrix $\tilde S$ is 
$\tilde S = PS \{ 1 + (1 - P) S\}^{-1}$. Thus, we may deal with $\tilde S$ as to those $S$-matrix which connects the physical components of $z$ amplitudes of the state only.

\section{Conclusions}
We have developed ADDPM solvable in four-dimensional space-time at lowest order of perturbative theory based on canonical quantisation. The interaction between DM and SM is mediated by non zero mass DP the origin of that is dictated by DM current and the derivatives of scalar dilaton fields. Because of DP fields fluctuations in medium, we find out that both DM and DP fields are observables only under the influence of stochastic (random) vector states. The indefinite metric emerged in Green's functions does not appear  in asymptotic expressions for physical observables.  If the dilaton is lighter than $2 m_{\chi}$, it can decay via a $\chi$-loop into a pair of Dark photons, yielding four leptons or neutrinos in the final state. The latter can clarify a question of CP violation in dark sector through the interference of 4-lepton channel. The model considered in this paper has a phenomenological consequence to probe DM through searching of dilatons and Dark photons in multi-lepton channels at hadron and $e^{+}e^{-}$ colliders.

\end{document}